\documentclass[11pt]{article}
\usepackage[a4paper, margin=1in]{geometry}
\usepackage[utf8]{inputenc}

\usepackage{titlesec}
\titlespacing*{\section}{0pt}{12pt}{0pt}
\titlespacing*{\subsection}{0pt}{12pt}{0pt}
\usepackage{amsmath, amssymb, amsfonts}
\usepackage{graphicx}
\usepackage[english]{babel}

\newcommand{\R}{\mathbb{R}}

\DeclareMathOperator*{\argmax}{argmax}

\begin{document}

\title{Inferring the Localization of White-Matter Tracts using Diffusion Driven Label Fusion}
\author{G Gallardo, G Zanitti, M Higger, S Bouix, D Wassermann}
\date{}

\maketitle

\section*{Abstract}
Inferring which pathways are affected by a brain lesion is key for both pre and post-treatment planning. However, many disruptive lesions cause changes in the tissue that interrupt tractography algorithms. In such cases, aggregating information from healthy subjects can provide a solution to inferring the affected pathways. In this paper, we introduce a novel label fusion technique that leverages diffusion information to locate brain pathways. Through simulations and experiments in publicly available data we show that our method is able to correctly reconstruct brain pathways, even if they are affected by a focal lesion.

\section{Introduction}
Pathologies such as infiltrative tumors or focal lesions disrupt the structure of white matter, resulting in cognitive deficits. Depending on the type and severity of the pathology, fiber bundles can be displaced, infiltrated or directly interrupted~\cite{Schonberg2006, Huisman2009, Won2016}. Inferring which pathways are closely located to the lesion or being directly affected by it is key for palliative care, as well as pre and post-treatment planning. With this knowledge, neurologists and neurosurgeons can decide if a lesion should be treated more aggressively or conservatively~\cite{Huisman2009, McGirt2009}. However, the most common method to reconstruct brain pathways, tractography, fails in the presence of brain lesions~\cite{Pictorial2004}. Therefore, having a non-tracking method to locate brain pathways which is robust to lesions is key to successfully treat patients with disruptive brain lesions.

In the presence of a disruptive brain pathology, the directionality commonly present in a Diffusion Weighted Image (DWI) can be severely reduced\cite{Pictorial2004}. This is directly reflected as a decrease in the fractional anisotropy (FA) in the affected regions. Since tractography algorithms rely on directionality information in order to reconstruct tracts, areas with significantly decreased FA will hamper or completely stop the tracking. This impedes reconstructing tracts in the brain, and therefore detecting those being affected by the lesion~\cite{Schonberg2006, Pictorial2004}. In such cases where tracking is not feasible, one possible solution is to rely on information from healthy subjects. Assuming we identified major bundles in a group of healthy subjects, we can register them to our patient’s brain and aggregate them to infer the affected pathways.

Label fusion is a family of techniques to infer a region's location in a target subject, based on its characteristics in a group of control subjects\cite{Asman2013}. One well known label fusion technique is Majority Voting~\cite{Xu1992}. Given a voxel on a brain image, each template subject "votes" for a label. The resulting label for the voxel will be that with the most votes. Majority Voting is simple to implement and has been shown to generate accurate brain segmentations even when using few template subjects~\cite{Asman2013}. Majority Voting weights the vote of each template subject equally, despite how (dis)similar they are from the target subject. However, among the templates there will be subjects more accurate in predicting a region's location, and those less accurate\cite{Rohlfing2004}. Because of this, it has been proposed to weigh the vote of each template based on some similarity function with the target~\cite{Sabuncu2010}. The underlying intuition is that the label choice should be driven by those subjects who resemble the most to the one being labeled. The practical advantages of various strategies based on this idea have been demonstrated by Artaechevarria et al.~\cite{Artaechevarria2009}.

\begin{figure*}[h]
    \includegraphics[width=\textwidth]{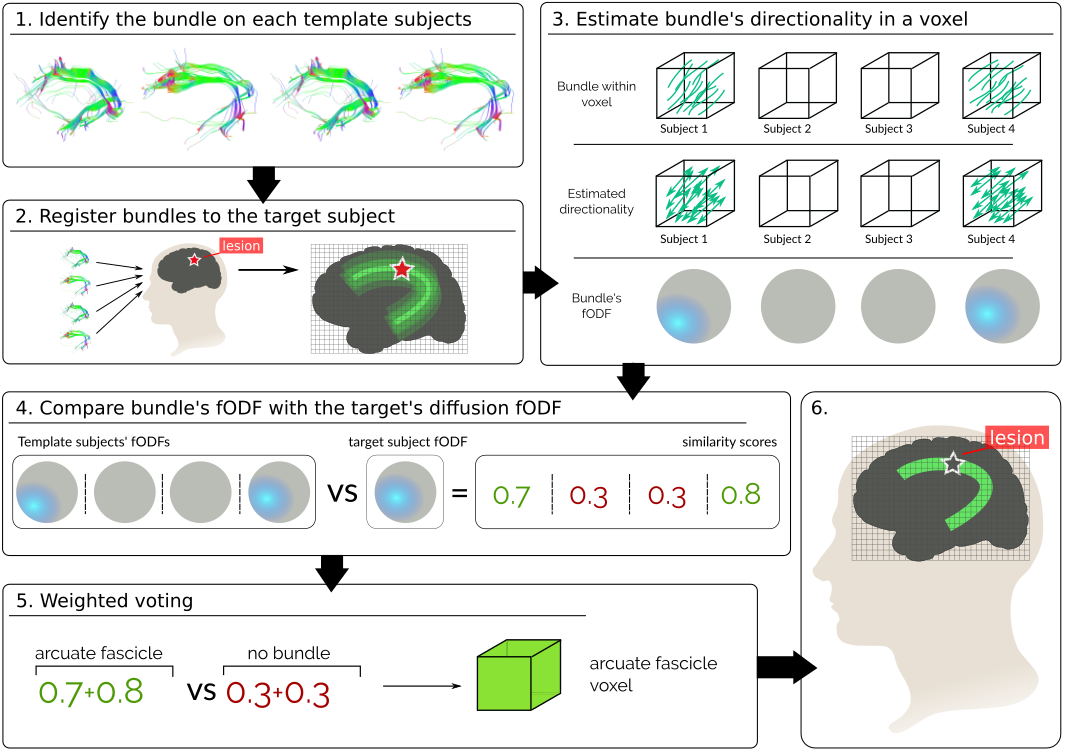}
    \caption{We use a label fusion technique to infer the location of white
             matter bundles by aggregating information of healthy subjects.
             After registering the tracts, each subject "votes" for either
             a tract or a non-tract structure (gray matter or background).
             Our technique adds diffusion based weights to each one of these votes.
             The weights are computed based on the similarity between the fiber
             Orientation Distribution Function of the structure being voted
             and the DWI of the target subject.}
    \label{fig:weighted_diffusion}
\end{figure*}

In this work, we introduce the first label fusion technique that leverages diffusion data, as represented in DWI, to weights the vote of each template subject. Since the presence of a path constrains the diffusion of water particles, adding diffusion information can help to better delineate fiber bundles~\cite{Girard2017}. Particularly, we have subjects voting for the presence of a specific tract in a region. Our technique then weighs the votes based on how much the voted pathway is supported by the target's diffusion data (Fig. \ref{fig:weighted_diffusion}). In this way, we give higher weights to the voted pathways which directionality better agrees with the target's diffusion.

We validate our technique through simulations and experiments in publicly available data. First, we validate that our technique correctly weights votes using synthetic data. For this we generate three types of synthetic DWIs (phantoms) representing: a single fiber phantom, a 90 degrees fiber crossing phantom, and a phantom with no fibers. We generate tracts from the DWIs and test how our technique weighs them and their planar rotations on each phantom. We show that our technique assigns higher weights to the votes for tracts aligned with the DWI's diffusion. After, we randomly select 50 subjects from the Human Connectome Project (HCP) and reconstruct 6 major brain pathways on each of them. We use the reconstructed tracts as ground truth to benchmark the inferences made by our technique and Majority Voting on a leave-one-out cross-validation experiment. Our results show that our proposed technique has a lower sensitivity than Majority Voting, but a higher specificity. Meaning, our technique obtains less false positives at the cost of obtaining more true negatives. Finally, we simulate focal lesions with different degrees of severity in the white matter of one HCP subject. Particularly, we target a spherical region where the Superior Longitudinal Fasciculus passes by. We do so by iteratively decreasing the FA of each voxel in the spherical region, until obtaining isotropic diffusion. We show that, while a tractography-based reconstruction technique fails to reconstruct, our technique is able to identify the affected tract. Furthermore, our technique is always able to correctly label the voxels surrounding the lesion, while labeling less voxels within the lesion as it worsen. This helps not only to detect the affected tract, but also to know how much it is affected.

\section{Methods}
\label{sec:methods}

\subsection{Majority Voting.}
Let $labels = \{l_i\}$ be the set of labels representing tracts in the brain. Let ${L_s}, s\in S$ represent the labeling of a set of template subjects $S$, where each $L_s \in labels^{v_x\times v_y \times v_z}$ is a 3D volume with dimension $(v_x,v_y,v_z)$ representing the labeling of template subject s. Majority Voting~\cite{Rohlfing2004} infers the label of each voxel $x$ in a target subject by computing:

\begin{equation}
\label{eq:mvoting}
\begin{aligned}
    L^*(x) = \argmax_{l \in labels} \sum_{s\in S} p(L(x) = l | L_s(x)),\\
    \text{where} \\
    p(L(x) = l | L_s(x)) =
    \begin{cases}
        1,& \text{if } L_s(x) = l \\
        0,& \text{otherwise.}
    \end{cases}
\end{aligned}
\end{equation}

In this case, each subject votes for a label, and the label with the most votes is assigned to the target voxel. It is important to notice that no information from the target subject is being used to infer the label $L^*(x)$.

\subsection{Diffusion Based Voting}
Majority Voting (Eq. \ref{eq:mvoting}) decides the label of a target voxel based on the ``votes'' of template subjects, without using any information from the target subject. However, when inferring the location of white matter pathways, we can profit from the fact that water molecules tend to diffuse in the direction of the pathways. This means that we can use diffusion information to weigh the voting process and help locate the bundles. In particular, votes for tracts aligned with the diffusion of the target subject should get higher weights.

One way to characterize the directionality on a diffusion weighted image (DWI) is by means of fiber orientation distribution functions (fODFs)~\cite{Tuch2004}. Now, we will first explain how to estimate a fODF from both DWI and tracts. Then, we will show how to leverage them in order to compute weights for each vote.

\subsubsection{Fiber Orientation Density Function from dMRI Data.}
By fitting the diffusion information into a Constrained Spherical Deconvolution (CSD) model, it is possible to estimate a fiber orientation density function\cite{Tournier2004} (fODF).  
The fiber ODF $F_x(\theta, \phi)$ represents the estimated fraction of fibers within the voxel $x$ that are aligned along the direction $(\theta, \phi)$, expressed in spherical coordinates.

\subsubsection{Fiber Orientation Density Function from Tractography.}
\label{sec:tract_fodf}
A tract can be described as a set of streamlines, where a streamline is a discretized 3-dimensional curve. Assuming that a streamline doesn't have sharp turns within a voxel, we can estimate its directionality by looking at the entry and exit points (Fig. \ref{fig:weighted_diffusion} A). Repeating this for each streamline on a tract, we obtain a set of directional vectors, representing the directionality of the tract within the voxel. As with the diffusion data, we can once more use the CSD model to estimate an fODF representing the directionality of a tract in the voxel.
   
\subsubsection{Label Fusion Weighted by Diffusion}
\label{sec:diffusion_fodf}
Majority Voting (Eq. \ref{eq:mvoting}) decides the label of a voxel based on how many subjects ``vote'' for it. Given that we are inferring brain pathways, we want to introduce a weight that denotes how much the voted tract resembles the target's diffusion data: 

\begin{equation}
\label{eq:mvoting_weighted}
L^*(x) = \argmax_{l \in labels} \sum_{s\in S} p(L(x) = l | L_s(x)) p(D(x) | D_{sl}(x)).
\end{equation}

In our segmentation scheme, the term $p(L(x) = l | L_s(x))$ is modeled as in the voting scheme (eq. \ref{eq:mvoting}). Our second term, $p(D(x) | D_{sl}(x))$ express the probability of seeing the diffusion of our target subject, $D(x)$, on voxel $x$, given the diffusion of subject $s$ generated by tract $l$ on the same voxel, $D_{sl}(x)$. Since registering DWIs is a highly time consuming task~\cite{ODonnell2017}, we want to avoid it. Instead, we can register tracts, for which efficient algorithms exist, and use them as an estimator of the diffusion
of each template subject. Knowing that water particles in the brain diffuse along tracts, we can estimate $D_{sl}(x)$ by computing the fODF of the registered tract $l$ as explained in \ref{sec:tract_fodf} . Simultaneously, we can characterize $D(x)$ with the fODF computed from the DWI of the target subject as explained in \ref{sec:diffusion_fodf}. In order to reflect how much the fODF of our target subject's diffusion resembles the fODF of the voted tract on a voxel, we model $p(D(x) | D_{sl}(x))$ as:

\begin{equation}
\label{eq:inner_odf}
\begin{aligned}
    p(D(x) | D_{sl}(x)) = 
    \begin{cases}
        \langle F(x), F_{sl}(x) \rangle,& \text{if } L_s(x) = l,\\
                        & \text{and } l \neq 0 \\
        \langle F(x), U \rangle,& \text{if } L_s(x) = 0 \\
        0,& \text{otherwise}
    \end{cases} \\
\end{aligned}
\end{equation}

where $F(x)$ is the fiber ODF on voxel $x$ estimated from the target's DWI, and is normalized such that $\langle F(x),F(x) \rangle = 1$. $F_{sl}(x)$ is the fiber ODF of the tract $l$ registered from subject $s$ and normalized as $F(x)$. $U$ is a uniformly distributed fiber ODF, this is the diffusion assumed for either the label 'no-tract', representing the background or a gray matter structure. By computing the inner product between normalized ODFs, we can estimate how much they look alike. In this way, we weight the vote of each subject accounting for
the white-matter structure of both the voting and target subjects.

\section{Experiments and Results}
In section \ref{sec:methods} we presented how to add diffusion information to Majority Voting (Eq. \ref{eq:mvoting_weighted}). This allows us to weigh the vote for a tract by how much the diffusion of the target supports it. Now, we validate our technique using both synthetic data and subjects from the Human Connectome Project (HCP). We start by assessing our techniques correctly weighs votes using DWI phantoms. Then, we proceed to infer the location of white-matter pathways in subjects of the HCP, and compare them with their ground-truth as reconstructed with the white-matter query language (WMQL)\cite{Wassermann2016}. Finally, we simulate a lesions in the white matter and test how our method behaves in its presence.

\subsection{Data and Preprocessing}

We created three types of diffusion weighted image phantoms using Phantomas~\cite{Caruyer2014}. The first phantom possess only one tract, traveling from one side to the other of the image horizontally (Fig. \ref{fig:pha_exp_1} A). The second possess two crossing tracts, forming a 90 degrees angle between them (Fig. \ref{fig:pha_exp_1} B). The last has no fibers on it, representing isotropic diffusion (Fig. \ref{fig:pha_exp_1} C). From each phantom we generated 30 diffusion weighted images (DWIs). All the DWIs were generated using a signal to noise ratio (SNR) of 20, and a resolution of $1mm$ per voxel. The final images are 3-dimensional matrices with 10 voxels in each dimension. Having such small images allows us to test our label fusion on a controlled environment. We estimated fiber orientation distributions (FODs) in each voxels by means of constrained spherical deconvolution (CSD)\cite{Tournier2007}  using Dipy~\cite{Garyfallidis2014}. The fODFs were discretized on a sphere with $n=100$ vertices. A uniform fODF, $U$, was created by assigning to each vertex of the sphere the value $\sqrt(n)/n$, making $\langle U, U\rangle = 1$.

To test our technique in more realistic scenarios, we randomly selected 50 subjects from the HCP500 dataset from the Human Connectome Project. For each subject, we first estimated FODs by means of CSD from they diffusion data. Then, we leveraged the FODs to perform whole-brain probabilistic tractography. For the tractography we used each voxel in the white-matter as a seed, and simulated 8 particles per seed~\cite{Garyfallidis2014}. Finally, we extracted four tracts from the left hemisphere using the implementation of the white-matter query language (WMQL)~\cite{Wassermann2016}. For each subject we computed non linear registrations to the rest using as reference their T1w images~\cite{Jenkinson2012}. Using the resulting warp transformations, we registered the tracts between every pair of subjects.

\subsection{Assessing the Correctness of Voting Weights in Synthetic Data}
\begin{figure*}[h]
    \includegraphics[width=\textwidth]{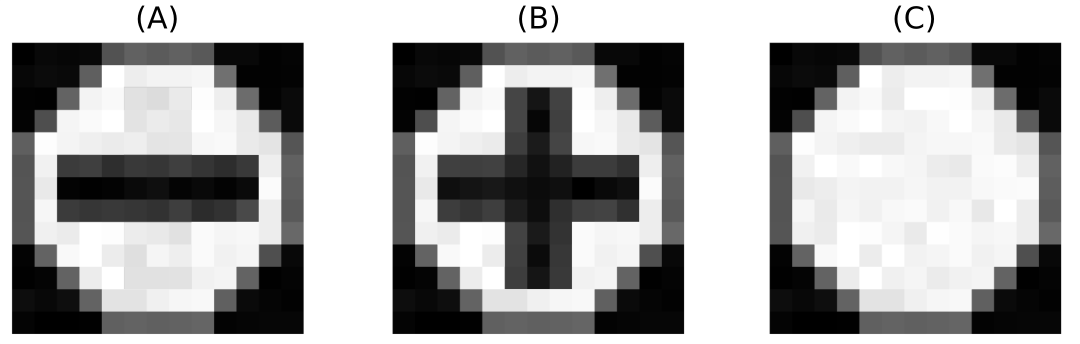}
    \caption{In order to test our technique, we created three types of synthetic DWIs, known as phantoms.
             (A) A phantom with only one tract in the white matter. (B) A phantom with two fibers crossing.
             (C) A phantom with no tracts, representing isotropic diffusion.}
    \label{fig:pha_exp_1}
\end{figure*}

\begin{figure*}[h]
    \includegraphics[width=\textwidth]{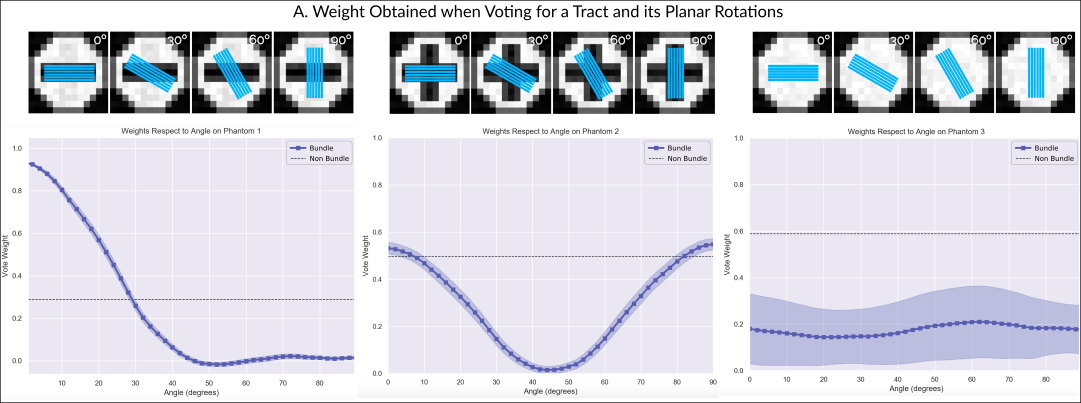}
    \caption{In order to study how a tract's directionality influences its vote weight,
             we estimated the fiber bundle by means of tractography in the Phantom A.
             Then, we computed the weights obtained by the tract and planar rotations of
             it in: (A) Phantom A. (B) Phantom B. (C) Phantom C. The figures show that
             our technique gives the highest weights to structures that are aligned
             with the underlying diffusion.}
    \label{fig:weights}
\end{figure*}

In order to study how the tract's directionality influences its vote weight, we started by reconstructing the tract present in the first phantom (Fig. \ref{fig:pha_exp_1} A). For this, we took one of the 30 generated DWIs, and computed 1000 streamlines by means of probabilistic tracking from the voxels through which the tract passes. If our technique behaves correctly, the tract we reconstructed should obtain a high vote weight in the DWIs derived from phantom A. At the same time, any change in its directionality, should decrease the received weight. To assess this, we computed weights in 30 DWIs for the reconstructed tract, and for planar rotations of it around the central voxel. Figure \ref{fig:weights} A shows the obtained weights on the first phantom. Effectively, the weight starts to rapidly decrease as the angle increments and the directionality of the tract moves away from that of the diffusion.

Figure \ref{fig:weights} B shows the weights obtained when computing weights of the reconstructed tract in 30 DWIs derived from the second phantom (Fig. \ref{fig:pha_exp_1} B), which have a fiber crossing in the central voxel. In this case, the weight is higher when the reconstructed tract aligns with one of the crossing fibers (at 0 degrees or 90 degrees), while rapidly decaying in between them. Finally, figure \ref{fig:weights} C shows the weights when using 30 DWIs with isotropic diffusion (Fig. \ref{fig:pha_exp_1} C). In this case, the weight is always low, driven by the discrepancy between the directionality of the tract and the free diffusion present in the DWIs.

To assess that the proposed model is not over-weighing tracts, we also computed the weight that a ``non-tract'' label would receive in each of the phantoms. This is, when the user votes that there is no tract present in the voxel. As explained in section \ref{sec:methods}, equation \ref{eq:inner_odf}, when a subject is voting for a non-tract label, a uniform fiber ODF is compared against the diffusion fODF. Figure \ref{fig:pha_exp_1} A, B, and C show the weight obtained in the central voxel when a subject is voting for the label ``non-tract''. In figure \ref{fig:pha_exp_1} A, we can see that the weight of ``non-tract'' is low, specially when compared with the high weight of the correctly aligned tracts (low angle rotations). This is driven by the highly directional underlying diffusion data of the phantom.  Figure \ref{fig:pha_exp_1} B shows that the weight of a ``non-tract'' vote is similar to that of an aligned tract. Finally, figure \ref{fig:pha_exp_1} C shows always a higher weight for the ``non-tract'' than for any tract, consistent with the isotropic diffusion of our third phantom.

\subsubsection{Inferring Tracts in Human Connectom Project Subjects}
To validate our technique in a more realistic but still controlled scenario, we used our technique to infer the location of specific white-matter tracts in 50 HCP subjects. We selected the following tracts to work with: Inferior Fronto-Occipital Fascicle (IFOF), Corticospinal Tract (CST), Inferior Longitudinal Fasciculus (ILF), and Superior Longitudinal Fasciculus (SLF I, II, and III). These 6 tracts provide a fair diversity of directionality, shape, and position in the brain.

For each tract we performed a leave-one-out cross-validation experiment. At each step, we inferred the tract of one subject from the registered tracts of the others using both Majority Voting and our technique. Then, we quantified the performance of both techniques using as ground truth the target's bundle. In particular, we computed their confusion matrix. A confusion matrix is a matrix $M \in \R^{labels \times labels}$, where each entry $M_{ij}$ represents the number of times the label in the ground truth was i and the technique labeled j. In this experiment, since we are only inferring one tract, we have two labels: tract and non-tract. We computed the sensitivity, and specificity of each confusion matrix~\cite{Kuhn2013}. Sensitivity measures the proportion of voxels in the ground-truth tract that were ``discovered''. Specificity measures the proportion of voxels that were correctly labeled, over all the labeled voxels. Table 1 shows the results obtained for each technique and tract. In all of the tracts, our technique shows on average a 5\% a lower sensitivity than Majority Voting. This means that we label a smaller portion of the ground-truth bundle. On the other hand, our diffusion weighted label-fusion achieves on average a 13\% higher specificity. Therefore, our technique is discovering less voxels, but those which are labeled can be trusted more.

\subsubsection{Inferring Tracts in the Presence of Simulated Lesions}
To test how our technique behaves on an injured brain, we simulated lesions at different degrees of severity in the white matter of one of our subjects. Given that some brain lesions directly affect Fractional Anisotropy (FA)~\cite{Schonberg2006, Huisman2009}, we simulated lesions by adding isotropic signal to a set of voxels, therefore lowering their FA. We targeted the SLF bundle, in order to compare how the labeling changes with lesion of different degrees. We did so by selecting a spherical region of $4mm$ where the SLF passes, and mixing the diffusion signal there with signal from the ventricles. Since the ventricles are regions filled with cerebrospinal fluid (CSF), their diffusion is approximately isotropic. In particular, for each voxel $x$ in the affected region, we chose a voxel $v$ in the ventricle and mixed their DWI signals, $S(\cdot)$, as follows:

\begin{equation}
    \label{eq:mixing}
S(x) = S(x)(1-\alpha) + S(v)\alpha, \alpha \in [0,1],
\end{equation}

where $\alpha$ manages the severity of the lesion. In this case, $\alpha=0$ represents healthy tissue, and $\alpha=1$ represents a total disruption of the white-matter, resulting in pure isotropic diffusion. 

We created four datasets by setting $\alpha$ to 0.25, 0.5, 0.75 and 1. On each dataset we performed a whole-brain probabilistic tractography, and filtered the resulting streamlines using WMQL to reconstruct the SLF. We found that for the values of $\alpha=0.75$ and $\alpha=1$ WMQL failed to recover the SLF, since it was interrupted by the simulated lesion. Then, we used both majority voting and our technique to aggregate the SLF from the rest of the subjects. Since Majority Voting for all datasets. Moreover, both Majority Voting and our technique obtained a high overlap in the result for $\alpha=0.25$. However, our technique labeled less voxels within the lesion as the alpha value increased (making FA to decrease). This is a good behaviour, since by lowering the FA we make the diffusion more isotropic, loosing the underlying tract. In particular, for $\alpha=1$, our technique results in two disconnected sets of voxels being labeled as SLF. This is correct, because for $\alpha=1$ the diffusion is completely isotropic, meaning that the tract was interrupted. This behaviour allows not only to know which tract was affected, but also at which degree it is affected, informing if the tract is interrupted or not.

\section{Discussion}
In this work we presented a label-fusion technique that leverages diffusion information to better infer the location of brain pathways. Our technique allows to correctly locate white-matter pathways even in the presence of a lesion that disrupts tractography algorithms. Furthermore, being based on a label-fusion algorithm, our technique can achieve accurate segmentations even when the inference is made from few subjects~\cite{Asman2013}. 

Our technique adds diffusion information in the process of label-fusion. Given that fiber bundles constrains the diffusion of water particles in the brain, our technique uses diffusion information to improve Majority Voting~\cite{Xu1992}. More specifically, we weight each vote based on how the voted pathway is supported by the target's diffusion data. In this way, voted pathways that better resemble the white matter of the target subject obtain a higher weight. The weights come from comparing how much the diffusion fODF of our target subject's resembles the fODF of the voted tract on a voxel. In this way, we can compare the white-matter structure of our target subject with that of the voting subject, without having to register DWIs. Adding diffusion weights to Majority Voting, allowed us to profit from its robustness while improving the labeling of white-matter bundles, as shown by our results in synthetic data and subjects from the Human Connectome Project.

\subsection{Our Technique Creates Weights Consistent With the Underlying Diffusion Data.}
To assess that our technique correctly weighed votes, we created three different phantoms, and derived DWIs from them. The phantoms had: (A) a single bundle running straight, (B) a 90 degrees crossing bundles, and (C) no bundles. We reconstructed the tract in the phantom A, and computed the weights obtained by the tract and planar rotations of it in the 3 phantoms. Our results show that our technique assigns high weights when the tract aligns with the bundles represented in the underlying diffusion data. Furthermore, when the directions differ by more than 10 degrees, the weight starts to drop rapidly. Our results also show that in the phantom C, where there is no bundle, the 'no-tract' vote receives a higher weight than any rotation. These results show that our technique is able to correctly weight each label based on diffusion directionality.

\subsection{Our Technique Shows Lower Sensitivity but Higher Specificity than Majority Voting.}
To test our technique in realistic data, we registered tracts between different subjects of the HCP. Using these registered tracts, we inferred the position of individual tracts in each subject. Table 1 shows that in each inferred tract, our technique achieved on average a 13\% higher specificity at the trade off of obtaining a 5\% lower sensitivity. Hence, our techniques presents less false positives at the cost of more false negatives, making it more conservative than Majority Voting. Furthermore, our specificity is always higher than 0.7. This means that our technique was able to discover less voxels belonging to the tracts, but at least 70\% of those labeled are correct.

\subsection{Our Technique Allows to Infers Which Tract is Affected by a Focal Lesion, and How Affected the Area Is}
We further characterized the behaviour of our technique by simulating non-deforming lesions in the white-matter of a HCP subject. Particularly, we defined a spherical region on the path of the Superior Longitudinal Fasciculus, and increased its Fractional Anisotropy until achieving isotropic diffusion. In doing so, we effectively simulated a focal lesions ranging from an infiltration to the interruption of a tract. Under these common scenarios in which the diffusion data is affected, reconstruction techniques such as WMQL are not able to recover the tract. This is because, as the diffusion data losses directionality, the tracking becomes more and more hampered. Indeed, in our experiment WMQL was not able to dissect the affect SLF, since the process of tractography was stopped by the lesion. On the other hand, both majority voting and our technique were able to correctly label the affected tract. However, as shown by our results (Figure \ref{fig:labeling}), only our technique was able to label less voxels as the tract became more interrupted. This is the expected behaviour in the presence of a lesion, since the more damaged an area is, the less it is expected for the tract to be present. When the tract is completely interrupted, our technique is still able to label the voxels surrounding the lesions, thus enabling to detect the affected tract. Therefore, our technique not only allows to detect the affected tracts, but also gives information on how altered the tract is.

\section{Conclusions}
\begin{figure*}[h]
    \includegraphics[width=\textwidth]{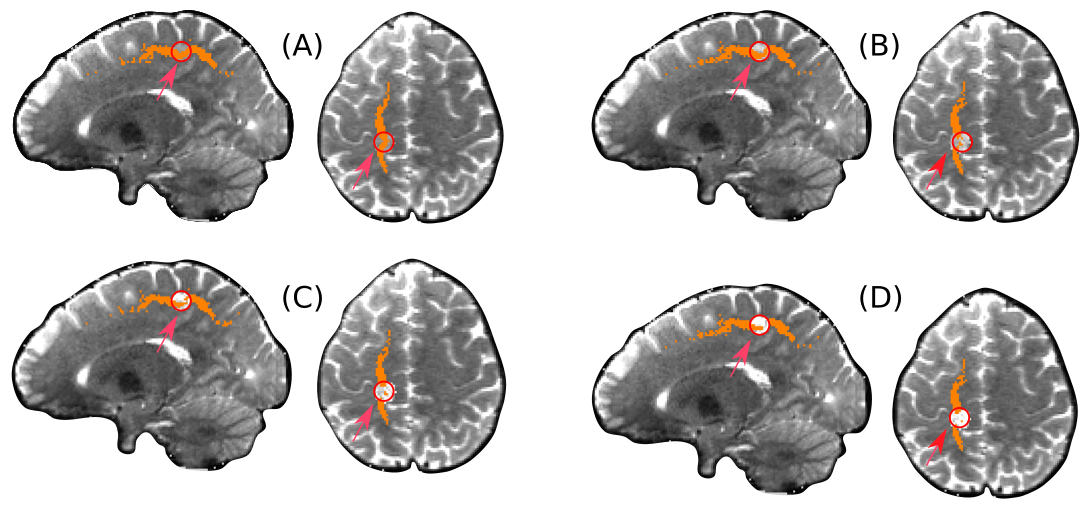}
    \caption{We tested how our technique behaves in the presence of simulated lesions.
             Lesions were simulated in a specific region (red circle) by following
             eq. \ref{eq:mixing} to lower the FA of the region. The SLF passes through
             such a region. The figures show the result of applying our technique 
             in order to determine the SLF at different values of the parameter $\alpha$ in
             eq. \ref{eq:mixing}: (A) $\alpha=0.2$, (B) $\alpha=0.5$, (C) $\alpha=0.75$, and (D) $\alpha=1$.
             Results show that while the value of $\alpha$ increases, the amount of voxels
             labeled within the affected region decreases.}
    \label{fig:labeling}
\end{figure*}

We presented a label fusion technique that leverages diffusion data to infer the localization of white-matter tracts. The results show that our technique is well suited to create subject specific segmentation of the  brain. Our method could be use in specificity medicine to interrogate diffuse pathologies like TBI, even in the presence of small lesions (as shown in fig. \ref{fig:labeling}). Our technique could be used in the study of infiltrative brain tumors, where registration is less of a problem, and in the future, could be extended to larger tumors, due to the growing advances in registration based on tumor growth models.

\newpage

\begin{table*}[h]
    \small
\label{table:sensitivity}
\centering
    \caption{Sensitivity and specificity of our proposed
             method (Weighted) and a non-weighted voted (Majority) when inferring single bundles from 9 subjects. The inferred bundles are: Superior Longitudinal Fasciculus (SLF) I, II and III, Inferior Longitudinal Fasciculus (ILF), Cortico Spinal Tract (CST), and Inferior Occipito-Frontal Fascicle (IOFF).}
\begin{tabular}{|l||c|c||c|c|} 
\hline  
& \multicolumn{2}{c|}{Sensitivity} & \multicolumn{2}{c||}{Specificity} \\  
\hline     
&  Weighted & Majority & Weighted & Majority  \\   
\hline 
IOFF & 0.10 (0.04) & \bf{0.13 (0.04)} & \bf{0.71} (0.15) & 0.55 (0.15) \\ 
CST & 0.38 (0.03) & \bf{0.42} (0.03) & \bf{0.80} (0.04) & 0.70 (0.04) \\ 
ILF & 0.22 (0.01) & \bf{0.27} (0.02) & \bf{0.77} (0.04) & 0.63 (0.04) \\ 
SLF I & 0.19 (0.01) & \bf{0.27} (0.01) & \bf{0.80} (0.04) & 0.66 (0.05) \\ 
SLF II & 0.23 (0.01) & \bf{0.31} (0.01) & \bf{0.82} (0.04) & 0.68 (0.04) \\ 
SLF III & 0.11 (0.01) & \bf{0.17} (0.02) & \bf{0.71} (0.04) & 0.58 (0.05) \\

\hline
\end{tabular}
\end{table*}

\newpage

\newpage

\newpage

\newpage

\newpage
\bibliographystyle{ieeetr}
\bibliography{bibliography}
\end{document}